\documentstyle[myjasa,epsfig,times]{article}

\begin{document}
\today
\title{Nonlinear tube waves in permeable formations: Difference
frequency generation}
\author{Yaroslav Tserkovnyak}
\affil{Harvard University, Lyman Laboratory of Physics, Cambridge, Massachusetts 02138}
\author{David Linton Johnson}
\affil{Schlumberger-Doll Research, Old Quarry Road, Ridgefield, Connecticut 06877-4108}

\begin{abstract}
We extend earlier work on nonlinear tube wave propagation in
permeable formations to study, analytically and numerically, the
generation and propagation of a difference frequency, $\Delta
\omega = \omega_1 - \omega_2$, due to an initial pulse consisting
of carrier frequencies $\omega_1$ and $\omega_2$.  Tube waves in
permeable formations have very significant linear
dispersion/attenuation, which is specifically addressed here.  We
find that the difference frequency is predicted to be rather
easily measurable with existing techniques and could yield useful
information about formation nonlinear properties.

\end{abstract}

\date{\today}

\begin{article}

\section{Introduction}

A tube wave is an acoustic normal mode in which the energy is
confined to the vicinity of a fluid-filled cylinder within an
elastic solid.  From a practical point of view it is generally the
dominant signal which appears in a typical borehole-logging
measurement and thus it is important in a variety of contexts in
the search for hydrocarbon sources.

One of these contexts lies in the fact that the tube wave may
couple to fluid flow within the rock formation if the latter is
permeable.  The linearized tube wave propagation in this regime
has been extensively studied both theoretically and experimentally
\cite{Winkler89} (see also \cite{Pampuri98} and references
therein). In the present article, we use a model of the tube wave
due to \cite{Liu97}. According to the model, the fluid in the
borehole is separated from the porous formation by an elastic
membrane (mudcake) of finite thickness. As a tube wave propagates,
the membrane flexes in and out of the pores, thus forcing the
fluid to flow through the formation.  This leads to the coupling
between the tube wave and the acoustic slow wave in the formation,
which in turn leads to attenuation and dispersion of the tube
waves.  In formations of moderate to large permeability, this
mechanism is the largest known source of attenuation/dispersion of
the tube wave and is the reason why it is specifically considered
in the present article.

Quite apart from this effect it is also known that sedimentary
rocks have very large coefficients of nonlinearity and so
\cite{Johnson94} developed a theory for nonlinear tube waves
neglecting the effects of the permeable formation.  Later,
\cite{Johnson99} combined this theory with the linearized theory
of \cite{Liu97} to describe a situation when the two effects are
simultaneously present.  As a numerical demonstration of the
theory, \cite{Johnson99} considered the propagation of a
narrow-banded (long duration) pulse consisting of a single carrier
frequency.  He showed that for realistic system parameters, the
main signal (the fundamental) quickly decays, but before
completely disappearing it generates a second harmonic and a
low-frequency band (the ``self-demodulated" pulse) both of which
are due to the nonlinearity of the problem.  The second harmonic
decays even faster than the carrier, with the result that the
self-demodulated pulse eventually dominates the entire signal at
large enough distances.

Because the second harmonic decays so fast, often it is
advantageous to determine nonlinear characteristics by using
pulses which consist of two different carrier frequencies,
$\omega_1$ and $\omega_2$. In addition to the second harmonics
(above) nonlinear effects lead to the generation of a signal
centered around the difference frequency $\Delta\omega = \omega_1
- \omega_2$.  This component may be reasonably energetic while at
the same time it is not attenuated as much as either second
harmonic, or even as much as either carrier frequency.  Thus, in
this article we are motivated to consider the propagation of two
narrow-banded pulses whose frequency separation $\Delta\omega$ is,
say, 10\% of the central frequency.  Moreover, because $\Delta
\omega$ is not that different from $\omega_{1,2}$, it is often
possible to measure its amplitude with the same acoustic
transducers as for the fundamental.

The organization of the article is as follows.  First, we review
the theory and derive analytical results for the nonlinear
propagation of a pulse consisting of two different frequencies. We
derive an approximation for the propagation of the entire signal
and we find an analytical form for the energy  of the band with
frequency $\Delta\omega$.  Next, we report results of numerical
calculations for a few different parameter sets and we show a good
agreement between our analytical and our numerical results.  In
the last section, we make a brief summary of our work.

\section{Theory}
\label{t}

The dispersion and attenuation of the linear tube wave propagation has been studied in \cite{Liu97}. To simplify our discussion we use an approximate form of the dispersion relation from \cite{Johnson99}:
\begin{equation}
k_z(\omega)=\omega\left[S_\infty+\tilde{\Theta}(\omega)\right] ,
\end{equation}
where $k_z$ is the wave vector along z-direction (the tube axis), $\omega$ is the angular frequency of the wave, $S_\infty$ is the slowness at infinite frequency and $\tilde{\Theta}$ is given by
\begin{equation}
\tilde{\Theta}(\omega)=\frac{\rho_f}{S_\infty b \left[W_{\rm mc}+W_p(\omega)\right]} .
\end{equation}
Here, $\rho_f$ is the density of the borehole fluid, $b$ is the
borehole radius, $W_{\rm mc}$ is the mudcake membrane stiffness
defined in \cite{Liu97}, and $W_p$ characterizes permeability
effects. $W_p$ depends on the borehole fluid viscosity, $\eta$,
formation permeability, $\kappa$, and the diffusivity of the slow
wave
\begin{equation}
C_D=\frac{\kappa K_f^*}{\eta\phi}
\end{equation}
through the equation
\begin{equation}
W_p(\omega)=-\frac{\eta C_D k_{\rm Sl}H_0^{(1)}(k_{\rm
Sl}b)}{\kappa H_1^{(1)} (k_{\rm Sl}b)} .
\end{equation}
Here, $k_{\rm Sl}=\sqrt{\imath\omega/C_D}$ is the wave vector of
the slow-compressional wave and $H_{0,1}^{(1)}$ are Hankel
functions.

Tube waves in nonlinear hyperelastic and impermeable formations
have been studied in \cite{Johnson94}. In \cite{Johnson99} the
effects of linear attenuation/dispersion and nonlinearity have
been combined to obtain an approximate equation of motion for tube
wave propagation in a realistic borehole. In the retarded time
frame $\tau=t-S_\infty z$ this equation is
\begin{equation}
\frac{\partial p(z,\tau)}{\partial z}+\frac{\partial
F(z,\tau)}{\partial \tau}- \frac{\beta S_\infty^3}{2
\rho_f}\frac{\partial p^2(z,\tau)}{\partial \tau}=0 , \label{ee}
\end{equation}
where $p$ is the pressure, $\beta$ is a dimensionless parameter
defined in \cite{Johnson99}.  The function $F(z, \tau)$ is most
simply related to the acoustic pressure in the Fourier transform
domain:
\begin{equation}
\tilde{F}(z,\omega)=\tilde{\Theta}(\omega)\tilde{p}(z,\omega) .
\end{equation}
After performing the Fourier transform of Eq.~(\ref{ee}), one obtains
\begin{equation}
\left(\frac{\partial}{\partial z}-\imath
q(\omega)\right)\tilde{p}(z,\omega)+ \frac{\imath\omega \beta
S_\infty^3}{2 \rho_f}\tilde{p^2}(z,\omega)=0 , \label{em}
\end{equation}
where
\begin{equation}
q(\omega)=\omega \tilde{\Theta}(\omega)
\end{equation}
is the reduced wave number. In Eq.~(\ref{em}), $\tilde{p^2}(z,\omega)$ is the Fourier transform of the square of $p(z,\tau)$ ({\it not} the square of the Fourier transform).

In \cite{Johnson99}, Eq.~(\ref{em}) has been used to study the
generation of the second harmonic as well as that of a
low-frequency self-demodulated signal in a situation in which
initially the pressure is a narrow band/long duration pulse
centered on the frequency, $\omega_1$.  In the present article we
consider a situation when two narrow-banded pulses are initially
present. As discussed in the Introduction, for practical relevance
we will take the difference in the pulse frequencies
$\Delta\omega$ to be 10\% of the central frequency
$\omega^\prime$. The initial signal is given by
\begin{equation}
p(z=0,\tau)=E_1(\tau)\sin[\omega_1\tau+\phi_1]+E_2(\tau)\sin[\omega_2\tau+\phi_2] .
\label{ic}
\end{equation}
Here, $E_i(\tau)$ are the envelope functions, $\omega_1\equiv\omega^\prime+\Delta\omega/2$ and $\omega_2\equiv\omega^\prime-\Delta\omega/2$. For the numerical demonstration of the next section, we take the same envelope functions as in \cite{Johnson99}:
\begin{equation}
E_1(\tau)=E_2(\tau)=\frac{1}{2}P_0\exp\left[-(\tau/T_W)^{10}\right] .
\label{is}
\end{equation}
By setting $\omega_1=\omega_2$ and $\phi_1=\phi_2$ in
Eq.~(\ref{ic}) one recovers the pulse considered in
\cite{Johnson99}. In all three parameter sets used for the
numerical calculations in the present article (see Table~I), we
take $\Delta\omega/\omega^\prime=0.1$ and $\omega^\prime\times
T_W=125\pi$ so that initially the different signals we consider
all ``look'' the same.

There are two characteristic distances relevant to the problem:
the decay length of the linearized theory, $Z_{\rm
att}=1/\gamma(\omega^\prime)$, where
\begin{equation}
\gamma(\omega)=\omega\Im\left[\tilde{\Theta}(\omega)\right] ,
\end{equation}
and the distance over which a shock front would develop, in the absence of attenuation. The latter is found in \cite{Hamilton98} to be
\begin{equation}
Z_{\rm shock}=\frac{\rho_f}{\beta S_\infty^3 \omega^\prime P_0} .
\end{equation}
The Gol'dberg number
\begin{equation}
\Gamma=Z_{\rm att}/Z_{\rm shock}
\end{equation}
measures the importance of nonlinear effects relative to the linear. In the three parameter sets (Table~I), the amplitude $P_0$ is chosen so that $\Gamma=0.21$, i.e. the nonlinear effects are significant, but overall pulse propagation is dominated by linear dispersion/attenuation.

For $\Gamma\ll 1$, the nonlinear effects can be ignored to a first
approximation.  Because, by assumption, the signal consists of two
narrow-band pulses one has the usual result of linear acoustics:
\begin{eqnarray}
p^\prime(z,\tau)&=&E_1(\tau-\Delta S_{g1}z){\rm e}^{-\gamma_1z}\sin\left[\omega_1(\tau-\Delta S_{p1}z)+\phi_1\right] \nonumber \\
&&+E_2(\tau-\Delta S_{g2}z){\rm e}^{-\gamma_2z}\sin\left[\omega_2(\tau-\Delta S_{p2}z)+\phi_2\right] ,
\label{se}
\end{eqnarray}
where $\Delta S_p$ ($\Delta S_g$) is the additional phase (group)
slowness relative to $S_\infty$:
\begin{eqnarray}
\Delta S_p(\omega)&=&\Re\left[\tilde{\Theta}(\omega)\right] , \nonumber\\
\Delta S_g(\omega)&=&\frac{\rm d}{{\rm d}\omega}\Re\left[\omega\tilde{\Theta}(\omega)\right] .
\end{eqnarray}
Throughout this paper, the subscripts 1 and 2 correspond to
frequencies $\omega_1$ and $\omega_2$, respectively. The physical
meaning of Eq.~(\ref{se}) is transparent: The pulse envelope
propagates with the group velocity and attenuates while each peak
and trough travels with the phase velocity. If the envelope is
broad enough, the relevant attenuation/dispersion quantities
should be evaluated at the central frequency of the pulse.

If the Gol'dberg number $\Gamma$ is small compared to unity but is
not completely negligible, Eq.~(\ref{se}) still gives a reasonable
approximation for the propagation of the original pulses but,
because of the quadratic nonlinearity in Eq.~(\ref{em}), new
frequency components will be generated which are centered around
$\omega=2\omega_1$, $2\omega_2$, $0$ as well as those centered
around $\omega_1-\omega_2$ and $\omega_1+\omega_2$.  Apart from a
direct numerical solution of Eq.(\ref{em}) one can develop a
perturbation theory thereof:

\begin{equation}
p(z, \tau) = p^\prime(z,\tau) + p_0(z, \tau) + p_\Delta(z, \tau) +
p_{2\omega_1}(z, \tau) + p_{2\omega_2}(z, \tau) + p_{\omega_1 +
\omega_2}(z, \tau) + \cdots \label{pert}
\end{equation}
Here, $p_0$ refers to the self-demodulated signal, $p_\Delta$
refers to the signal whose bandwidth is centered around $\Delta =
\omega_1 - \omega_2$, etc.  By substitution of Eqs. (\ref{se}) and
(\ref{pert}) into Eq.(\ref{em}) one can derive an approximate
equation for the evolution of these nonlinearly generated
components, even in the presence of significant linear dispersion
and attenuation.

In this manner the generation of the second harmonic and of the
self-demodulated signal has been studied in \cite{Johnson99}. In
the present article we focus on the component centered around the
frequency $\Delta=\omega_1-\omega_2$. (The analytical expressions
we derive for this mode, can easily be generalized to the
$\omega=\omega_1+\omega_2$ case by redefining
$\omega_2\rightarrow-\omega_2$.)  We obtain
($\Delta\phi\equiv\phi_1-\phi_2$)
\begin{equation}
p_\Delta(z,\tau)=E_1(\tau-\Delta S_{g1}z)E_2(\tau-\Delta S_{g2}z)\frac{\Delta\omega\beta S_\infty^3}{2\rho_f}\times\Re\left[{\rm e}^{-\imath(\Delta\omega\tau+\Delta\phi)}\frac{{\rm e}^{\imath(q_1-q_2^*)z}-{\rm e}^{\imath q(\Delta\omega)z}}{q(\Delta\omega)-q_1+q_2^*}\right] .
\label{pd}
\end{equation}
One is reminded that this equation is valid only for narrow-banded pulses ($\Delta\omega\times T_W\gg 1$) and small nonlinear effects ($\Gamma<1$).

Next, we consider the energy of the carrier
($\omega\approx\omega^\prime$) band, ${\mathcal{E}}^\prime(z)$ as
well as that of the band centered on the difference frequency
$\omega\approx\Delta\omega$, ${\mathcal{E}}_\Delta(z)$. We define
them by
\begin{eqnarray}
{\mathcal{E}}^\prime(z)&\equiv&\int_{\omega^\prime/2}^{3\omega^\prime/2}
\left|\tilde{p}(z,\omega)\right|^2{\rm d}\omega , \label{en1}\\
{\mathcal{E}}_\Delta(z)&\equiv&\int_{\Delta\omega/2}^{3\Delta\omega/2}\left|\tilde{p}(z,\omega)\right|^2{\rm
d}\omega . \label{en2}
\end{eqnarray}
Using Parseval's theorem and Eqs. (\ref{se}) and (\ref{pd}), one
gets
\begin{eqnarray}
{\mathcal{E}}^\prime(z)/\pi&\approx&\int_{-\infty}^{+\infty}\left|\tilde{p}^\prime(z,\omega)\right|^2{\rm d}\omega \nonumber \\
&=&{\rm e}^{-2\gamma_1z}\int_{-\infty}^{+\infty}E_1^2(\tau){\rm d}\tau+{\rm e}^{-2\gamma_2z}\int_{-\infty}^{+\infty}E_2^2(\tau){\rm d}\tau
\label{ep}
\end{eqnarray}
and
\begin{eqnarray}
{\mathcal{E}}_\Delta(z)/\pi&\approx&\int_{-\infty}^{+\infty}\left|\tilde{p}_\Delta(z,\omega)\right|^2{\rm d}\omega \nonumber \\
&=&\left(\frac{\Delta\omega\beta S_\infty^3}{2\rho_f}\right)^2\left|\frac{{\rm e}^{\imath(q_1-q_2^*)z}-{\rm e}^{\imath q(\Delta\omega)z}}{q(\Delta\omega)-q_1+q_2^*}\right|^2\times\int_{-\infty}^{+\infty}E_1^2(\tau-\Delta S_{g1}z)E_2^2(\tau-\Delta S_{g2}z){\rm d}\tau .
\label{ed}
\end{eqnarray}

In the special case that the two envelope functions are identical,
$E_1(\tau)\equiv E_2(\tau)$, one can derive a simple analytic
result for the nonlinearly generated signals centered around
$\Delta\omega$ as well as the self-demodulated signal centered
around $\omega =0$.  For this, we rewrite the input signal
(\ref{ic}) in the form
\begin{equation}
p(z=0,\tau)=E^\prime(\tau)\sin[\omega^\prime\tau+\phi^\prime]\nonumber ,
\label{ip}
\end{equation}
where $\phi^\prime\equiv(\phi_1+\phi_2)/2$ and
\begin{equation}
E^\prime(\tau)=2E_1(\tau)\cos\left[(\Delta\omega/2)\tau+\Delta\phi/2\right] .
\label{im}
\end{equation}
$E^\prime(\tau) $ is now viewed as a narrow-banded envelope
function for the $\omega^\prime$ mode (assuming that
$\Delta\omega\ll\omega^\prime$).  As before, the total signal can
be approximated by
\begin{equation}
p(z,\tau)\approx p_{[slow]}(z,\tau)+p^\prime(z,\tau)  \label{ss}
\end{equation}
where $p_{[slow]}(z,\tau) \equiv p_0(z,\tau) + p_\Delta(z,\tau)$
now includes both the self-demodulated component as well as the
components centered around $\Delta\omega$.  Within the context of
the foregoing approximations it is given by
\begin{equation}
p_{[slow]}(z,\tau)=\frac{\beta
S_\infty^3}{4\rho_f}\int_{-\infty}^{+\infty} \frac{-\imath\omega
\tilde{E^{\prime 2}}(\omega)}{\imath\omega\Delta S_g^\prime-
2\gamma^\prime-\imath q(\omega)}\times\left[{\rm
e}^{(\imath\omega\Delta S_g^\prime- 2\gamma^\prime)z}-{\rm
e}^{\imath q(\omega)z}\right]{\rm e}^{-\imath\omega\tau}{\rm
d}\omega . \label{so}
\end{equation}
The primed quantities $\Delta S_g^\prime$ and $\gamma^\prime$ are
evaluated at the frequency $\omega^\prime$.

\section{Numerical Results}
\label{r}

For the numerical calculations, we consider three parameter sets,
\{A, B, C\}, which are listed in Table~I. These parameters are
identical to those considered in \cite{Johnson99}, where the
relevance of these parameters to realistic borehole properties is
discussed.  There is one change in that in this article we take
$T_W$ to be five times larger.  This is done in order to make the
band widths of the pulses so narrow that the $\Delta\omega$ mode
clearly separates from the rest of the low frequency signal.

\begin{figure*}
\center{\includegraphics[scale=0.7]{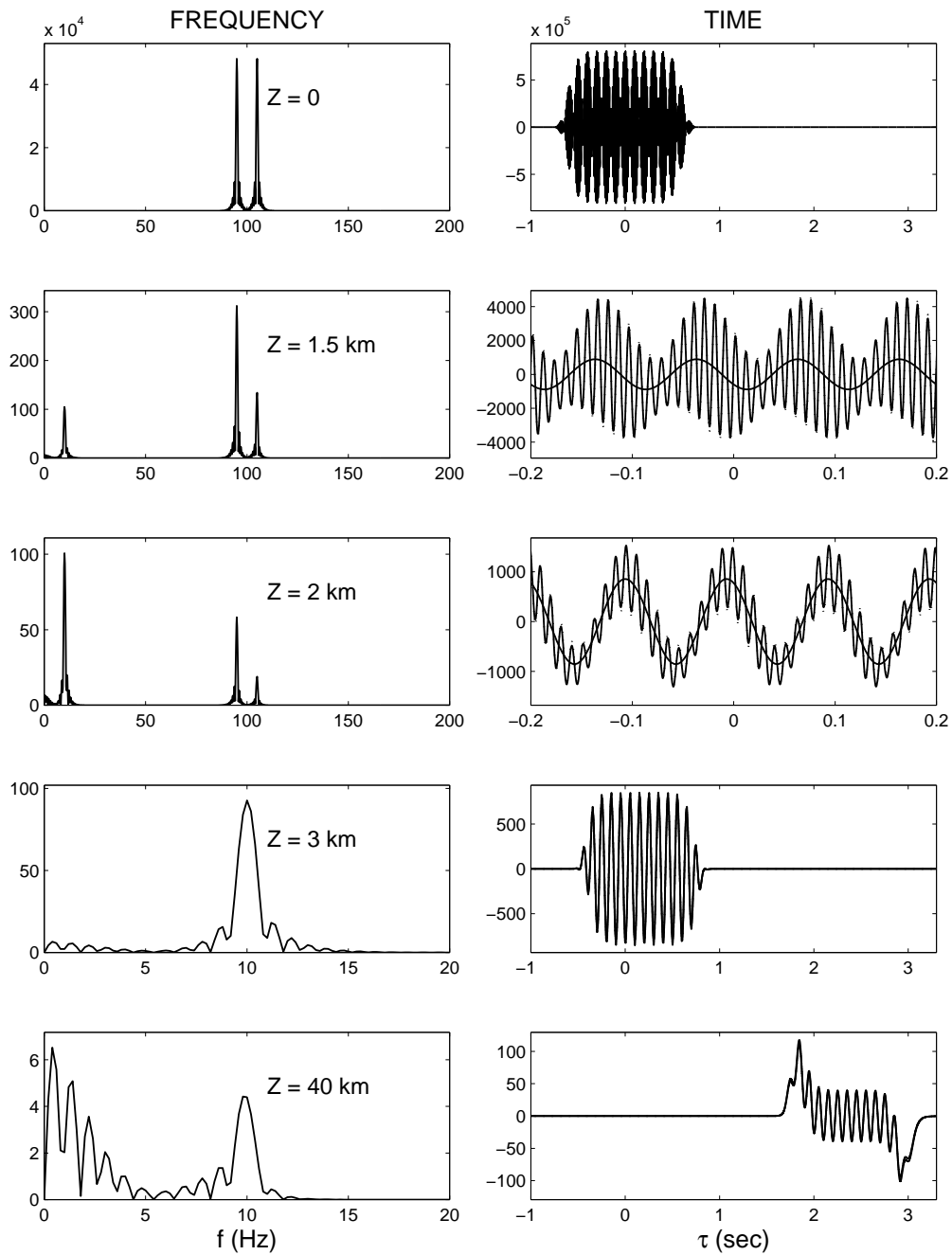}} \caption{Calculated
pulse evolution using parameter set A, from Table~I. The right
column is the signal in the time domain, the left is the frequency
spectrum. Also shown as a solid line on the right is the
analytical expression for the low-frequency pulse, Eq.~(\ref{so}),
toward which the pulse evolves. Similarly, the analytical
expression for the total signal, Eq.~(\ref{ss}), is shown as a
dotted line: It essentially overlies the numerically calculated
signal.  Notice the various changes of scale.} \label{a0}
\end{figure*}

\begin{figure*}
\center{\includegraphics[scale=0.7]{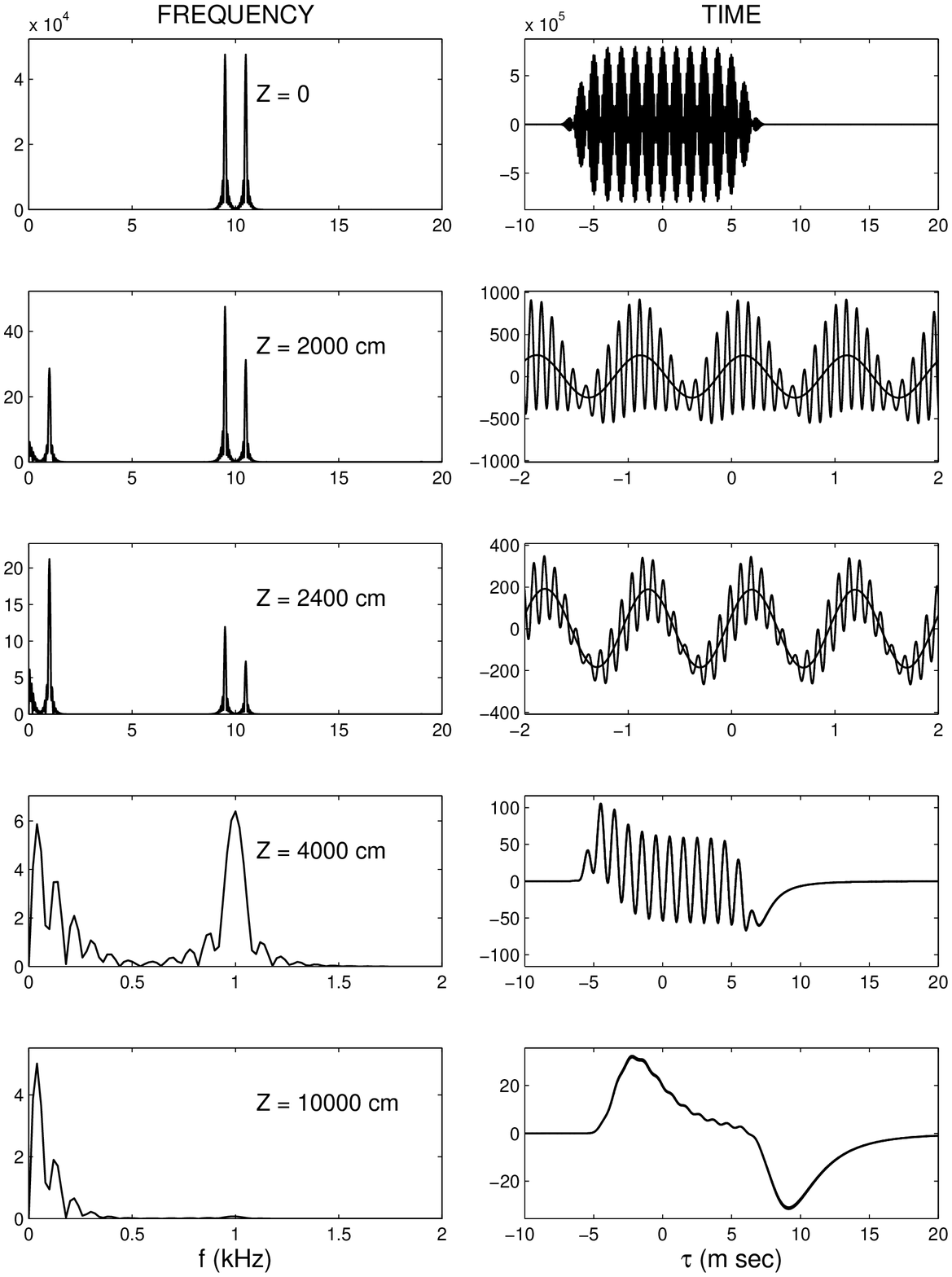}}
\caption{Calculated pulse evolution using parameter set B, from Table~I. Same conventions as Fig.~\ref{a0}.}
\label{b0}
\end{figure*}

\begin{figure*}
\center{\includegraphics[scale=0.7]{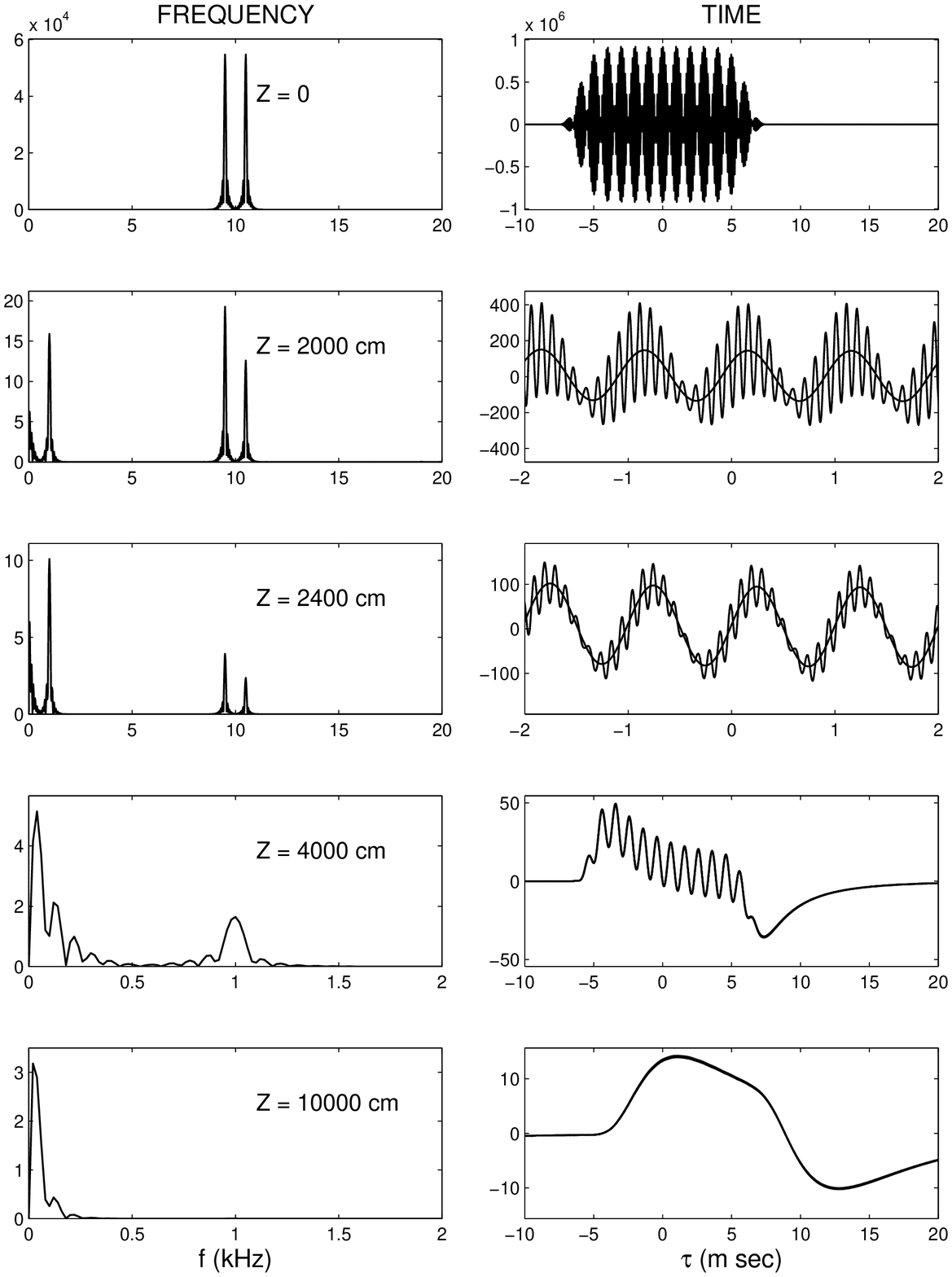}}
\caption{Calculated pulse evolution using parameter set C, from Table~I. Same conventions as Fig.~\ref{a0}.}
\label{c0}
\end{figure*}

In order to solve the equation of motion (\ref{em}) we use the
same Lax-Wendroff algorithm as in \cite{Johnson99} and the initial
pulse (\ref{ic}) with envelope functions (\ref{is}) and phases
$\phi_1=\phi_2=0$. Figs.~(\ref{a0}), (\ref{b0}), and (\ref{c0})
show several snapshots of the pulse profile in both time and
frequency domains for three parameter sets (Table~I) A, B and C,
respectively. Cases B and C have similar initial pulses, but
differ somewhat in their dispersion relations, as can be seen from
the corresponding parameter sets in Table~I. Sample A has a center
frequency two orders of magnitude lower than samples B and C, and
as a consequence, the relevant modes have much longer attenuation
length $Z_{\rm att}$.

It is pedagogically useful to first examine Fig.~\ref{c0} in
detail. Considering the left column of plots, one can trace an
intuitively clear sequence of events: Initially, the signal is
given by two sharp pulses separated by 10\% frequency difference
$\Delta\omega$ (first row of plots). As this signal propagates
through the borehole and gradually attenuates, modes centered at
$\omega\approx\Delta\omega$ are being generated (second row). They
soon start to dominate the general shape of the pulse because they
attenuate less than does the carrier (third row). Concomitantly,
$\omega\approx0$ modes start to appear and give a significant
``background'' for the $\Delta\omega$ signal (fourth row). In the
end, all higher-frequency components decay and the completely
self-demodulated signal $\omega\approx0$ contains most of the
energy (fifth row). One has to be careful looking at the right
column of the plots: The black profile of the first plot outlines
an envelope Eq. (\ref{im}) of the 10kHz signal, detailed structure
of which can be seen only if we expand the time scale $\tau$
(second and third plots). In order to examine the
$\Delta\omega\approx100$Hz components and the lower-frequency
signal, we again show the entire signal in the last two plots.
Although Eq. (\ref{ss}) is also plotted, it is indistinguishable
from the results of the full numerical calculation.  For large
enough distance $p(z, \tau)$ evolves to $p_{[slow]}(z,\tau)$, Eq.
(\ref{so}), as is indicated in the plot.

Qualitatively Fig.~\ref{a0} and Fig.~\ref{b0} look similar to
Fig.~\ref{c0}. But in the case of sample A, because of the smaller
attenuation, the $\Delta\omega$ component does not demodulate
completely even after propagating a distance of $\sim40$km.

\begin{figure*}
\center{\includegraphics[scale=0.9]{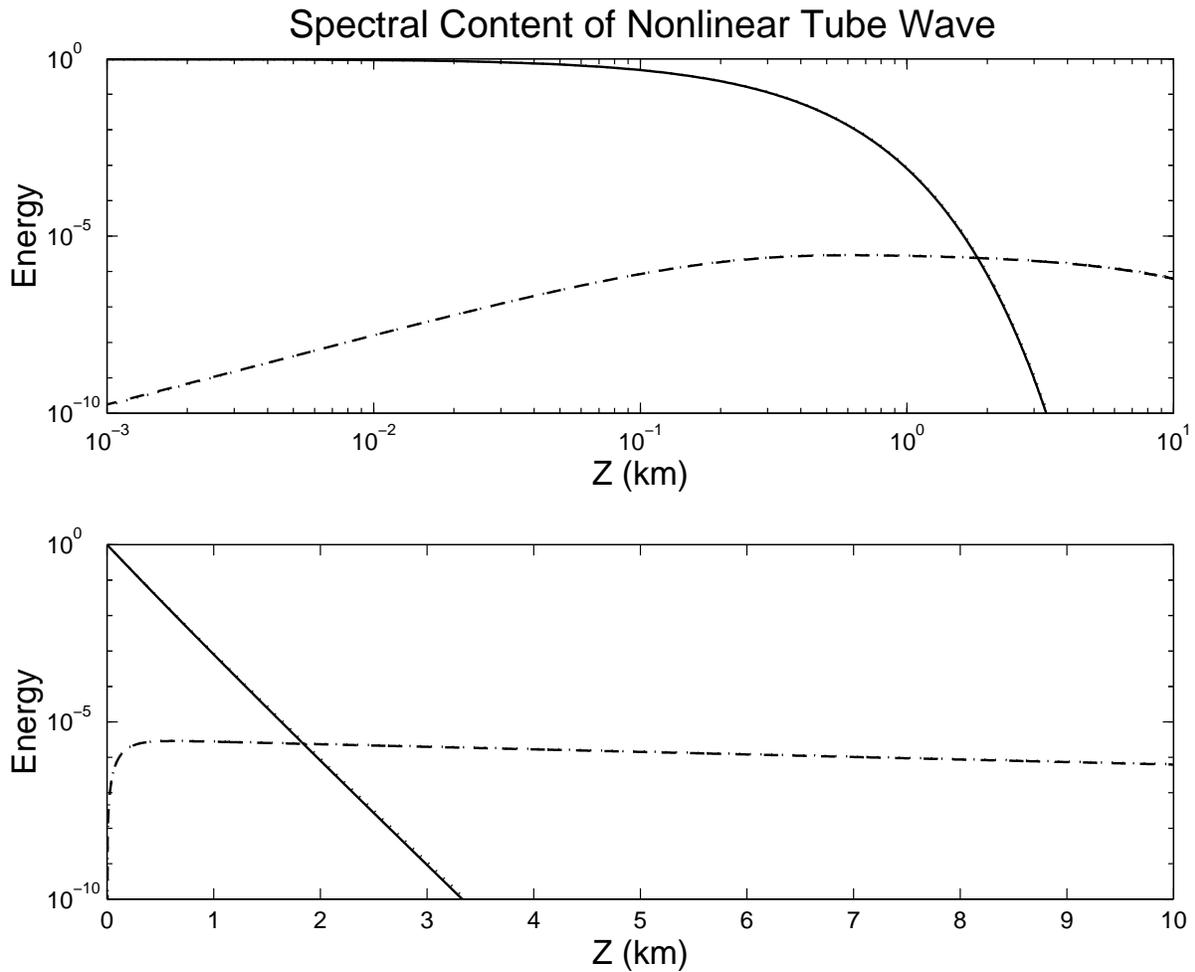}} \caption{Numerically
calculated values of the energy ${\mathcal{E}}^\prime(z)$ (solid
line) and ${\mathcal{E}}_\Delta(z)$ (dashed line) defined by
Eqs.~(\ref{en1}) and (\ref{en2}) for the waveforms of
Fig.~\ref{a0}, sample A. Each curve has been normalized by the
value of ${\mathcal{E}}^\prime(z=0)$. The data are shown in both
log-log and semilog plots, in order to emphasize the short and
long distance behavior, respectively. The dotted lines show the
analytical expressions, Eqs.~(\ref{ep}) and (\ref{ed}); the fact
that they are almost indistinguishable from the numerically
calculated curves is precisely the point.} \label{a1}
\end{figure*}

\begin{figure*}
\center{\includegraphics[scale=0.9]{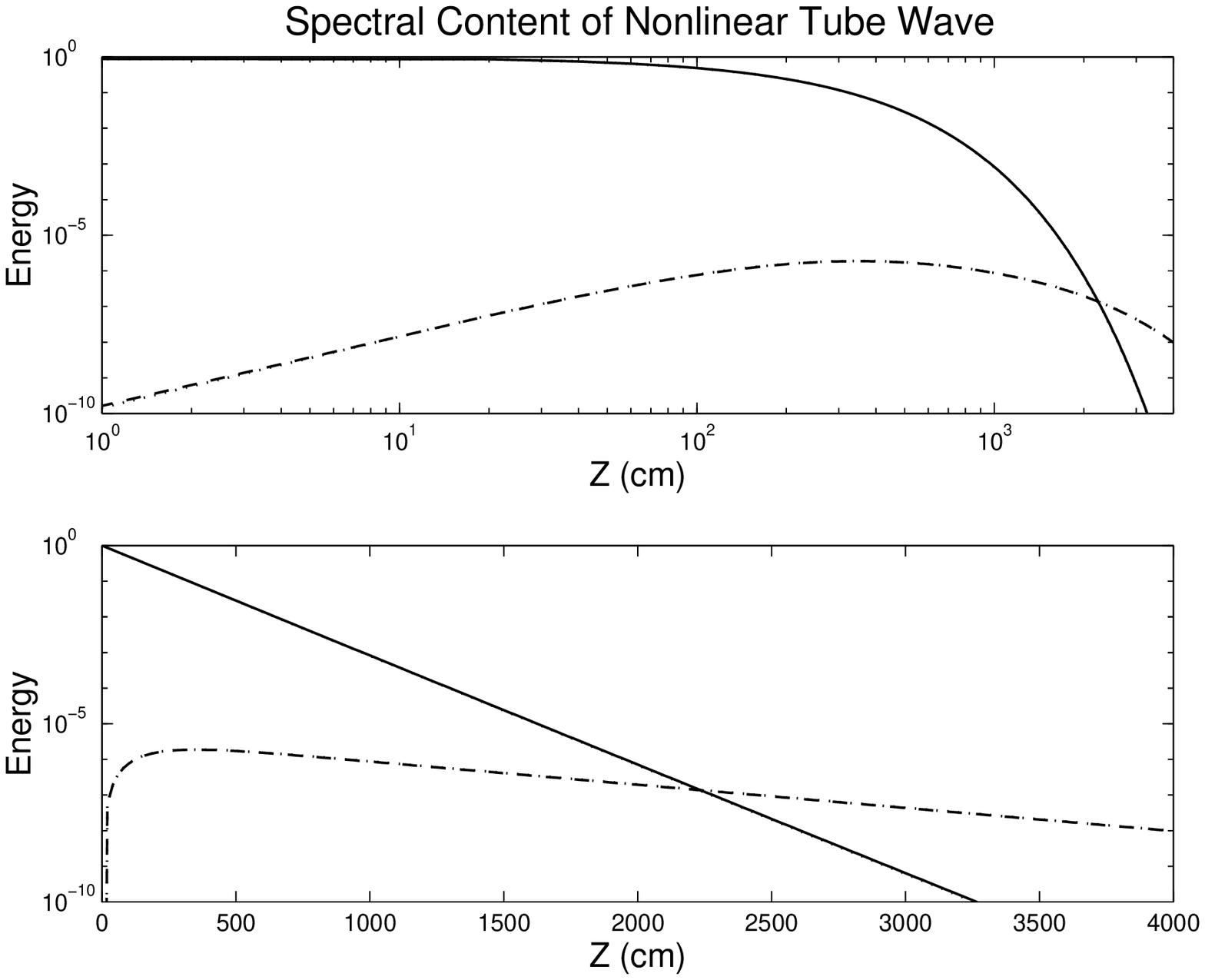}}
\caption{Numerically calculated values of the energy ${\mathcal{E}}^\prime(z)$ and ${\mathcal{E}}_\Delta(z)$ for the waveforms of Fig.~\ref{b0}, sample B. Same conventions as Fig.~\ref{a1}.}
\label{b1}
\end{figure*}

\begin{figure*}
\center{\includegraphics[scale=0.9]{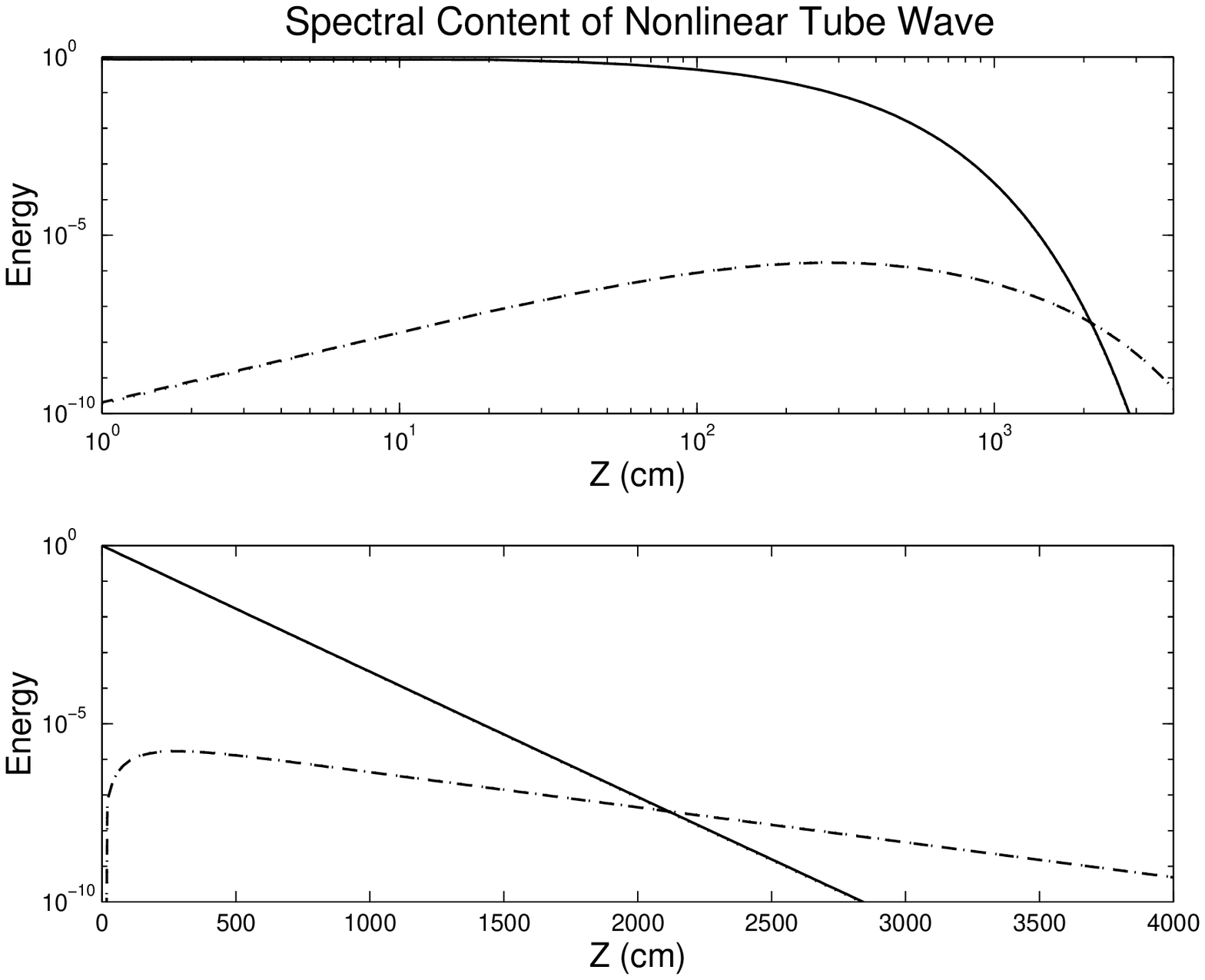}}
\caption{Numerically calculated values of the energy ${\mathcal{E}}^\prime(z)$ and ${\mathcal{E}}_\Delta(z)$ for the waveforms of Fig.~\ref{c0}, sample C. Same conventions as Fig.~\ref{a1}.}
\label{c1}
\end{figure*}

Next, for each of the three cases, A, B, and C, we calculate the
energy of the $\omega=\omega^\prime$ band directly using
Eq.~(\ref{en1}) and also using Eq.~(\ref{ep}).  Similarly, we
calculate the energy of the $\omega=\Delta\omega$ band using
Eq.~(\ref{en2}) as well as Eq.~(\ref{ed}).  The results are
plotted in Figs.~\ref{a1}, \ref{b1} and \ref{c1}.  One can compare
Fig.~\ref{b1} with Fig.~6 of \cite{Johnson99} which shows the
energy of the second harmonic and the $\omega=0$ band for an
almost monochromatic initial pulse and same system parameters. As
can be intuitively expected, the $\Delta\omega$ band is
intermediate in both the maximal energy it gains and the distance
it propagates, as compared to the lower-frequency band and the
higher harmonics. The $\Delta\omega$ component does not reach
energies as high as the second harmonics do, but it propagates
much further, still carrying a significant fraction of energy; it
is $\sim55$dB down from the initial carrier energy,
${\mathcal{E}}^\prime(z=0)$, in our examples. On the other hand,
while the lower-frequency signal persists longer than the
$\Delta\omega$ modes, it is never as energetic and it would be
more difficult to measure even in principle.

The intersections of the solid and dashed curves in
Figs.~\ref{a1}, \ref{b1}, and \ref{c1} indicate the crossover
region, where the $\omega=\Delta\omega$ band starts dominating
over the carrier band $\omega=\omega^\prime$, as can be seen in
Figs. \ref{a0}, \ref{b0}, and \ref{c0}, respectively.

\section{Conclusions}
\label{c}

We have used the theory of \cite{Johnson99} of tube wave
propagation in permeable formations to describe nonlinear
interaction of two narrow-banded pulses. The theory incorporates
both nonlinear effects and a realistic model for
dispersion/attenuation of tube waves. We have extended this
previous work on the propagation of a single narrow-banded pulse
to describe the generation and propagation of the
$\omega=\Delta\omega$ band when two different carrier frequencies
are present in the initial pulse.

We have derived analytical results for the self-demodulated
component, the $\Delta\omega$ band, and the total signal in the
regime of weak nonlinearity and they are in excellent agreement
with an accurate numerical calculation using three different
parameter sets.  Specifically, we have studied the spectral
content of the signal and demonstrated that the $\Delta\omega$
band can have a potential application because of its long
attenuation length and its relatively high energy content. Also,
if $\Delta\omega/\omega^\prime\sim 0.1$, as we consider here,
there is a practical bonus as the same transducers used for the
generation of the carrier signal can, presumably, be used for the
detection of the $\Delta\omega$ band.

\begin{acknowledgments}
This work was supported in part by the NSF grant DMR 99-81283.
\end{acknowledgments}

\appendix

\end{article}

\begin{table}
\caption{Values of input parameters for the calculation of tube wave characteristics.}
\begin{tabular}{llcccl}
\hline
\hline
Sample & & A & B & C & \\
\hline
& $\phi$ & 0.30 & 0.30 & 0.30 & \\
& $\eta$ & 0.01 & 0.01 & 0.01 & poise \\
& $K_f^*$ & 2.25 & 2.25 & 2.25 & GPa \\
& $b$ & 0.1 & 0.1 & 0.1 &m \\
& $\rho_f$ & 1000. & 1000. & 1000. & kg/m$^3$ \\
Input & $S_\infty$ & 667. & 667. & 667. & $\mu$sec/m \\
parameters & $\beta$ & 50.5 & 50.5 & 50.5 & \\
& $W_{\rm mc}$ & 250. & 250. & 100. & GPa/m \\
& $\kappa$ & 2. & 0.2 & 0.2 & $\mu$m$^2$ \\
& $\omega^\prime/2\pi$ & 0.1 & 10. & 10. & kHz \\
& $T_W$ & 625. & 6.25 & 6.25 & msec \\
& $P_0$ & 81. & 80. & 92. & kPa \\
\\
\hline
\\
& $C$ & 15. & 1.5 & 1.5 & m$^2$/sec \\
& $Z_{\rm att}$ & 278. & 2.8 & 2.4 & m \\
Calculated & $Z_{\rm shock}$ & 1323. & 13.4 & 11.6 & m \\
quantities & $\Gamma=Z_{\rm att}/Z_{\rm shock}$ & 0.21 & 0.21 & 0.21 & \\
& $\Delta S_p(\omega^\prime)$ & 55.9 & 6.8 & 6.9 & $\mu$sec/m \\
& $\Delta S_g(\omega^\prime)$ & 52.7 & 3.5 & 3.5 & $\mu$sec/m \\
\hline
\hline
\end{tabular}
\label{T1}
\end{table}

\end{document}